\documentclass[twocolumn,footinbib,pre,showpacs,showkeys]{revtex4}

\usepackage{graphicx}

\begin{document}

\title{Fractal analysis of Sampled Profiles: Systematic Study}
\date{\today}
\author{C. Castelnovo, A. Podest{\`a}, P. Piseri, P. Milani}
\email{pmilani@mi.infn.it}
\homepage{http://webcesid1.fisica.unimi.it/~labmilani/} \affiliation{INFM
- Dipartimento di Fisica, Universit{\`a} degli Studi di Milano,\\ Via
Celoria 16, 20133 Milano, Italy}

\begin{abstract}
A quantitative evaluation of the influence of sampling on the numerical
fractal analysis of experimental profiles is of critical importance.
Although this aspect has been widely recognized, a systematic analysis of
the sampling influence is still lacking. Here we present the results of a
systematic analysis of synthetic self-affine profiles in order to clarify
the consequences of the application of a poor sampling (up to 1000 points)
typical of Scanning Probe Microscopy for the characterization of real
interfaces and surfaces. We interprete our results in term of a deviation
and a dispersion of the measured exponent with respect to the ``true''
one. Both the deviation and the dispersion have always been disregarded in
the experimental literature, and this can be very misleading if results
obtained from poorly sampled images are presented. We provide reasonable
arguments to assess the universality of these effects and we propose an
empirical method to take them into account. We show that it is possible to
correct the deviation of the measured Hurst exponent from the ``true''one
and give a reasonable estimate of the dispersion error. The last estimate
is particularly important in the experimental results since it is an
intrinsic error that depends only on the number of sampling points and can
easily overwhelm the statistical error. Finally, we test our empirical
method calculating the Hurst exponent for the well-known 1+1 dimensional
directed percolation profiles, with a 512-point sampling.
\end{abstract}


\maketitle

\section{Introduction}
The characterization of interfaces and of the mechanisms underlying their
formation and evolution is a subject of paramount importance for a broad
variety of phenomena such as crystal growth, rock fracture, biological
growth, vapor deposition, surface erosion by ion sputtering, cluster
assembling, etc ... (\cite{bar95,man82,fam84,mea98,pietro85} and
references therein). Since the pioneering work of B.B. Mandelbrot, fractal
geometry has been widely used as a model to describe these physical
systems that are too disordered to be studied with other mathematical
tools but that still hold a sort of ``order'' in a scale-invariance sense
\cite{bar95,man82,vic92}. In particular, the growth of interfaces
resulting from the irreversible addition of subunits from outside (vapor
deposition of thin films, low energy cluster beam deposition, etc...)
shows a typical asymmetric scale invariance, because of the existence of a
privileged direction (e.g. the direction of growth)
\cite{kar86,wit83,fam85,mes85,mea86,mea87,mea98,lic86,fam90,kri95,csa92,bal90,mul90,sit99,yoo99,mil01a,mil01b,buz00}.
These interfaces belong to the class of self-affine fractals and they can
be described either by the fractal dimension $D$ or by the well-known
Hurst exponent $H$ \cite{pei89,mea89,vos89,sch95a,man85,man86}. If these
systems are the result of a temporally evolving process, they usually show
also a time scale-invariance described by the exponent $\beta$
\cite{vic92,bar95}. Because of the close relationship between the scaling
exponent(s) and the fundamental mechanisms leading to scale invariance,
universality classes can be defined \cite{vic92,bar95}. An accurate
knowledge of $H$ (and $\beta$) is required to identify the universality
class of the system and to give a deep insight on the underlying formation
processes.

The possibility of characterizing the topography of an interface in a
dimension range from the nanometer up to several tens of microns, in a
relatively simple and quick way by Atomic Force Microscopy (AFM) and
Scanning Tunneling Microscopy (STM) \cite{bin82,bin86} has stimulated an
upsurge of experimental report claiming for self-affine structures (see
Refs. \cite{mal97,mal98} and references therein). The abundance of
experimental characterization of different systems and the limited
sampling capability of the scanning probe microscopies (SPM) prompted at
the attention of many authors the need of an accurate methodological
approach to the determination of the exponent $H$ and of its error
\cite{den99a,tat98}, realistically considering the consequences of the
finite sampling inherent to SPM. Typical sampling with an AFM or a STM is
256 or 512 points per line, for a maximum of 512 lines. Most of the
results published in the late eighties and early nineties were based upon
256x256-point data-sheets, or even smaller ones (see list of references in
Ref. \cite{mal97}). Commercially available SPMs offer today a maximum of
512x512-point resolution, and homemade instruments hardly go beyond this
value.

Many authors have questioned the reliability of the measurement of the
Hurst exponent from a poorly sampled profile
\cite{dub89,sch95,meh97,alm96}. In order to quantify the influence of the
sampling on the determination of $H$, a numerical analysis can be
performed on artificial self-affine profiles, generated with a specific
algorithm, with a fixed number of points $L$ and known Hurst exponent
$H_{in}$. The ``true'' exponents ($H_{in}$) are then compared with the
ones measured directly from the generated profiles ($H_{out}$). Usually a
sensible discrepancy between the measured $H_{out}$ and the expected
$H_{in}$ is found \cite{dub89,meh97,alm96}. The discrepancy is not uniform
but depends on the value of $H_{in}$. As one would expect, the discrepancy
is globally dependent on the number $L$ and it approaches zero for large
values of $L$. In particular, for $L<1000$ the sampling effect is of great
importance since the discrepancy can be of the order of the exponent
itself (100\% relative error) \cite{sch95}. Dubuc \textit{et al.} have
reported that even for values of $L$ as high as 16384, the discrepancy is
still significant \cite{dub89}.

Although the problem of sampling has been clearly addressed and discussed,
quite surprisingly a systematic analysis of the problem, considering
different generation algorithms, is still lacking. The dependence of the
sampling effect on $L$ has been investigated \cite{dub89,sch95} and also
many different methods for the measurement of $H_{out}$ have been
considered for different values of $H_{in}$ in the range $[0.1-1]$
\cite{dub89,sch95,meh97,alm96}. However, either only one single generation
algorithm has been used \cite{sch95,alm96}, or the results from different
generation algorithms have not been compared \cite{meh97}. We believe that
this comparison is of fundamental importance.

Indeed profiles from different generation algorithms can be considered as
different self-affine objects sampled in $L$ points. For a fixed value of
$H_{in}$, these objects would all have the same fractal dimension if they
were sampled with an infinite number of points. The fundamental question
at this point is whether the discrepancy of $H_{out}$ from $H_{in}$, for a
finite value of $L$, is the same for every self-affine object (i.e. for
every generation algorithm). Only an analysis that considers different
self-affine objects has a statistical validity and allows a reliable
interpretation of the results. Up to now the results obtained in
literature from a single generation algorithm did non allow a discussion
of the nature of the aforementioned discrepancy, which has been
interpreted as an uncontrollable error affecting the analysis of sampled
profiles. The main conclusion drawn by these authors is the
non-reliability of results obtained from profiles with less than 1024
sampling points \cite{sch95}.

Our aim is to achieve a deeper understanding of the effects of sampling in
order to answer the question whether the measurement of the Hurst exponent
with a poor number of sampling points is reliable or not. This point is
crucial both for future analysis of self-affine profiles and for a correct
interpretation of the results already present in literature.

From a more general point of view, fractality is characterized by the
repetition of somehow similar structures at all length scales and can be
described in its major properties by a single number: the fractal
dimension $D$ \cite{fal90,man82}. Any finite sampling of a fractal object
poses both an upper and a lower cut-off to this scale invariance. It has
been shown that these cut-offs introduce a deviation in $D$ and the
sampled object has a dimension different from the one of the underlying
continuous object \cite{dub89,alm96,meh97}. However, it is still unknown
whether the sampling influences in a different way different objects
characterized by the same ideal dimension, thus breaking the sort of
universality that makes a fractal be identified by its dimension only.

In this paper we present a systematic analysis considering together all
the generation algorithms found in literature. The aim of our analysis is
to understand whether the discrepancy of the measured $H_{out}$ for a
fixed $L$ and for every generation algorithm is completely random or has a
universal dependence on $H_{in}$. The latter observation can be
interpreted as a reminiscence of the fact that a fractal object is
completely characterized by its dimension \footnote{As pointed out by Voss
\textit{et al.} in Refs. \cite{vos86a,vos86b} this is not completely true,
because fractal objects have also properties that do not depend on the
fractal dimension only, such as lacunarity. However, the main statistical
properties of a fractal object are characterized by its dimension
\cite{man82,fal90}}. The distinction is of crucial importance because in
the case of universal dependence of $H_{out}$ on $H_{in}$ , one can
empirically correct the discrepancy of the measured exponents from the
``true'' ones. Some authors independently suggested to use directly the
$H_{out}$ vs. $H_{in}$ curves as correction, but they considered only one
generation algorithm without discussing the universal character that these
curves must have in order to be utilized for any self-affine object
\cite{den99a}.

Conversely, on the basis of our analysis, we will interpret the
discrepancy in terms of two distinct contributions: a universal deviation
and a random dispersion. We will propose a powerful method to correct the
universal deviation and we will discuss the nature of the dispersion,
which is due to both statistical fluctuations and an intrinsic sampling
effect. The latter turns out to be a sort of systematic error that cannot
be corrected unless one knows the generation algorithm that produced the
self-affine object. In the case of generic self-affine profiles which have
not been generated by a specific algorithm, such as experimental profiles,
the above arguments no longer hold. A new procedure to quantify the
intrinsic error in the measurement of the Hurst exponent of generic
self-affine profiles is thus needed.

On these basis, we will discuss the effect of sampling on the reliability
of the fractal analysis of poorly sampled self-affine profiles, focusing
on both the deviation and the dispersion of the measured exponents from
the ideal ones, showing that the conclusions drawn by Schmittbuhl
\textit{et al.} that ``\textit{... a system size less than 1024 can hardly
be studied seriously, unless one has some independent way of assessing the
self-affine character of the profiles and very large statistical
sampling}'' were too restrictive \cite{sch95}. Moreover, we will point out
that the estimate of the intrinsic error is essential for a correct
classification of a process in terms of universality classes. In fact, in
order to distinguish exponents belonging to different classes, it is
necessary to quantify the error on the measurement. Up to now, the
statistical error or the error of the fit have been used to quantify the
error on the measurement of $H$ \cite{kri93,den99b,iwa93}. Both the
statistical error and the error of the linear fit can be made very small,
if a large number of profiles are averaged. However, if the measurement is
likely to be affected by more subtle intrinsic errors, such as the
aforementioned dispersion due to the sampling, considering only the
statistical error may lead to serious misleading. The intrinsic error in
many cases may indeed be much larger than the statistical one.

In the following sections we will present a systematic analysis of
synthetic self-affine profiles with the aim of both achieving a deep
understanding of the effects of sampling and providing the
experimentalists of a reliable tool for the fractal analysis of surfaces
and interfaces. To this purpose we have developed a new automated fitting
protocol in order to avoid any arbitrariness in the measurement. With this
new methodology we will study the effects of sampling, enlightening the
main characteristics of the deviation and the dispersion of the measured
exponents. We will present a new powerful method to correct the deviation
of $H_{out}$ and to estimate the error of the measurement. Finally, we
will apply our empirical correction procedure to 512-point profiles
created with the directed percolation (DP) algorithm \cite{bul92}. This
system provides a simple benchmark to test our protocol and allows
noticing the opportunity of the correction.

\section{The Automated Fitting Protocol}

Self-affine systems occurring in nature are usually profiles or surfaces.
In order to measure their Hurst exponents the 2+1 dimensional case of
surfaces is usually reduced to 1+1 dimensions, considering the
intersection of the surface with a normal plane. The particular case of
in-plane anisotropy results in a dependence of $H$ on the orientation of
the plane with respect to the surface \cite {dub89,sch95,fal90,bar95}.

Once we have scaled down the analysis to 1+1 dimensions, the following
general properties characterize a self-affine profile. If $h(x)$ is the
height of the profile in the position $x$, the orthogonal anisotropy can
be expressed by the scaling relationship:
\begin{equation}
h(\lambda x)=\lambda^H h(x)
\end{equation}
where $H \in (0,1)$ is the Hurst exponent, $\lambda$ is a positive scaling
factor and the equation holds in a statistical sense \cite{bar95,sta71}.
The fractal dimension $D$ of the profile is related to the Hurst exponent
by the equation $D=2-H$ while the dimension of the surface is $D=3-H$
\cite{man86,mor94}. The lower is $H$, the more space invasive is the
surface. In most of the physical self-affine surfaces, the scale
invariance does not extend to all length scales but there is an upper
cut-off above which the surface is no longer correlated. The length at
which this cut-off appears is defined as the correlation length $\xi$
\cite{bar95,mal97}. In the present analysis, we consider only profiles
whose correlation length (expressed in number of points) is equal to their
length $L$. To this purpose we have carefully studied each generation
algorithm in order to grant the condition $\xi = L$. For this reason we
were often forced to generate very long profiles and to consider only
their central portion \cite{mak96,meh97,sim99}. The usual procedure to
measure the Hurst exponent of a self-affine profile $h(x)$ is to calculate
appropriate statistical functions from the whole profile. These functions
of analysis (AFs) show a typical power law behavior on self-affine
profiles:
\begin{equation}
AF[h(\cdot),k]=c\,k^{f(H)}
\end{equation}
where $c$ is a constant, $k$ is a variable indicating the resolution at
which the profile $h$ is analyzed (typically a frequency or a
spatial/temporal separation), and $f(H)$ is a simple function of the Hurst
exponent $H$ \cite{meh97,bar95,yan97,mor94,pen94,pre86,sim98}. The power
law behavior of the AF is then fitted in a log-log plot in order to
calculate the exponent $H$. In the analysis of statistical self-affine
profiles there are random fluctuations super-imposed to this power law
behavior. The signal-to-noise ratio of these fluctuations is
scale-dependent, the AFs being calculated as averages of statistical
quantities at different length scales \cite{bar95}. To reduce this noise,
the average of the AFs obtained from $N$ independent profiles is usually
taken before the execution of the linear fit. However, while small-scale
fluctuations are easily smoothed, larger scale fluctuations converge very
slowly.

\begin{figure}
\begin{center}
\includegraphics{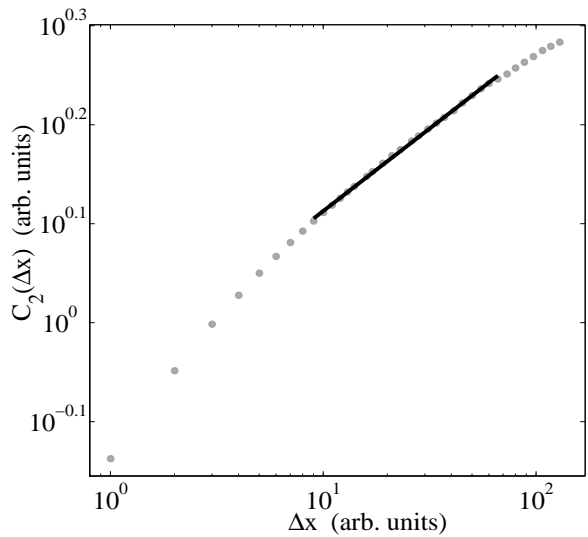}
\caption{\label{fig:1}Average height-height correlation function $C_{2}$
calculated from $N=500$ profiles of $L = 512$ points, generated with the
random addition method with Hurst exponent $H_{in} = 0.1$. It is also
shown the linear region and the fit obtained with the automated fitting
protocol (AFP). One can clearly see the overall curved shape due to the
sampling.}
\end{center}
\end{figure}

The identification of the linear region in the analysis of the AFs is a
puzzling point. Windowing saturation is present at length scales
comparable with the profile length depending on the nature of the profiles
\cite{yan97}. This results in a departure from the power law behavior to a
constant value. Moreover, the degradation of the fractality due to the
sampling causes a diversion of the AFs from their ideal power law
behavior. This produces both a discrepancy of the measured Hurst exponent
from the ideal value (a change of the slope in the log-log plot) and a
shortening of the linear region as shown in Fig. \ref{fig:1}.
Here, the presence of curved regions is clearly visible. It can be seen
that this anomalous behavior is not localized at length scales close to
the length of the profile, but involves also the shortest length scales
especially for values of $H$ close to zero. It is important to notice that
this effect is not due to experimental conditions, such as the finite size
of the SPM scanning probe. Thus it is necessary, in particular for small
values of $H$, to chose a linear region instead of fitting the whole
function. The methods proposed in the literature to identify the linear
region (e.g. the consecutive slopes method \cite{bou91,bar95}, correlation
index method \cite{bar89}, the coefficient of determination method
\cite{ham96} and the ``fractal measure'' method \cite{li99}) are usually
based on an arbitrary (human) choice. This is particularly delicate since
the curvature in the AFs can be so small, if compared to the statistical
noise, that it is hard to distinguish the correct linear region. Because
of this reason, we think that the proposed methods suffer of a high degree
of arbitrariness. Moreover, all these methods make no distinction between
a straight line with statistical noise and a slightly curved line.

Due to the previous arguments and since no universally accepted fitting
procedure is available in literature, we were prompted to develop an
automated fitting protocol (AFP) with two purposes: to reduce as much as
possible the effects of the curved regions on the measured exponent, and
to define a standard algorithm for the choice of the linear region,
eliminating, as much as possible, any arbitrariness. This is very
important for the reliability of the results, in particular for the
comparison of different generation algorithms. Moreover, the automation of
the fitting procedure is essential to perform a systematic analysis. In
fact, in order to have good statistics, a large number of AFs must be
calculated and fitted.

\begin{figure}
\begin{center}
\includegraphics{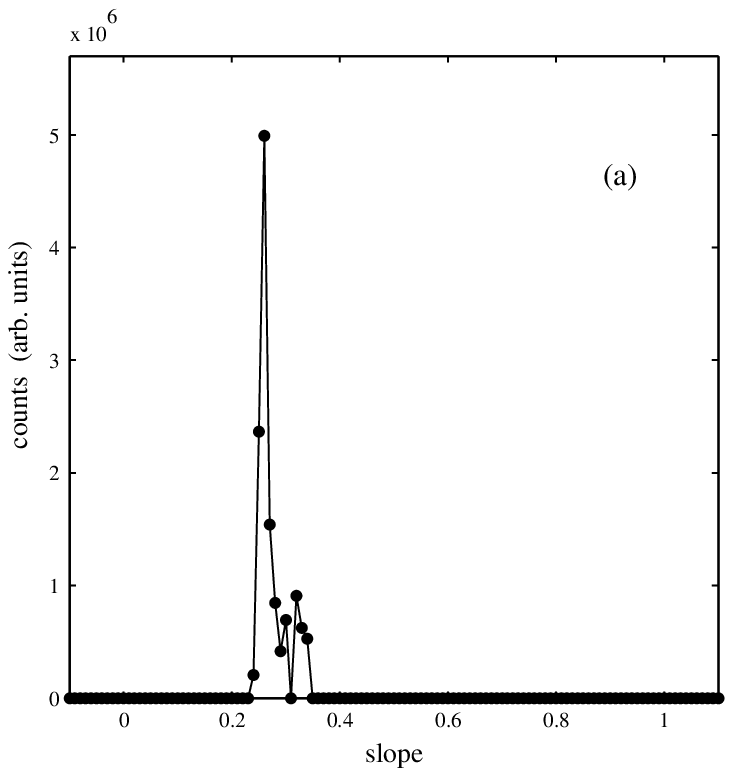}
\includegraphics{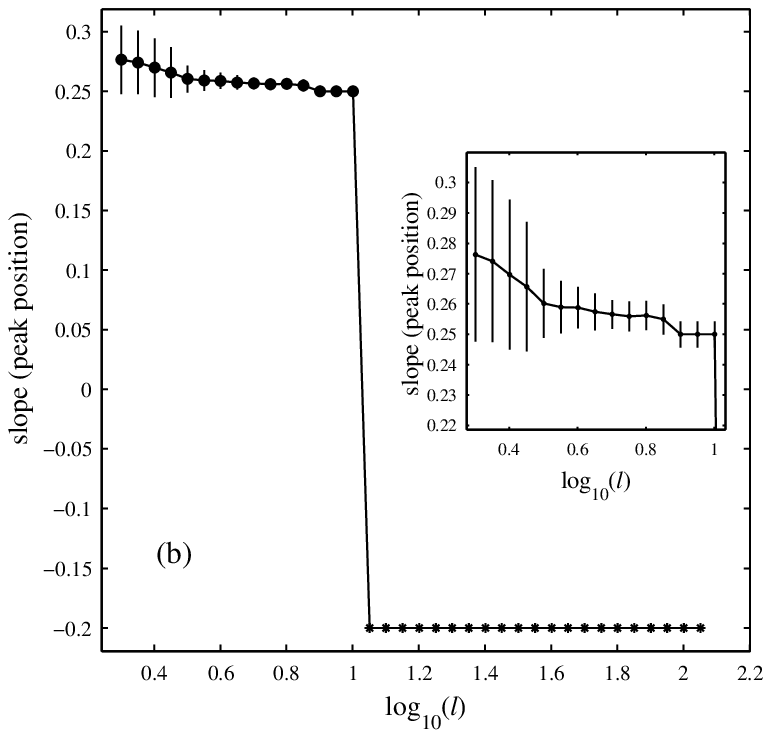}
\caption{\label{fig:2}Application of the fitting protocol step by step:
(a) the distribution of the slopes for a single value of the length $\ell$
of the portion ($\ell = 0.35$ decades) and (b) the final plot of the
slopes (peak positions) vs. $\ell$, with an inset magnification showing
the error bars.}
\end{center}
\end{figure}

In our procedure, that is an implementation of the consecutive slopes
algorithm \cite{bar95}, the curve to be fitted is divided in many portions
of the same length $\ell$ (in number of points) and each of them is
considered separately. A linear and a cubic fit are performed on each
portion. Comparing the mean distance of the linear fit from the portion to
the mean distance of the cubic from the linear fit, we evaluate whether
the portion is almost linear with uncorrelated noise or it presents a
definite curvature. Obviously, the distinction is not immediate and we
have to set a threshold to separate the two cases through a parameter in
the fitting procedure. The use of a parameter is common to other methods
(see for example the coefficient of determination method used in Ref.
\cite{ham96}). Once the fitting parameter is set, our procedure is able to
decide automatically whether the portion is ``curved'' or ``linear''. Only
the ``linear'' portions are then considered. They undergo a
straight-line-fit analysis through which the slopes and their errors are
determined. A distribution of the slopes weighted with the values of the
errors is then built (see Fig. \ref{fig:2}a) and its main peak position
and width are measured. We do not consider here the presence of more than
one linear region with different slopes. Thus, there is a well-defined
main peak in the distribution. We have extended our procedure also to the
case of more than one linear region, but this extension is out of the
scopes of this article.

The procedure described above is repeated varying the length $\ell$ of the
portions from a minimum value $\ell_{min}$  up to the length of the curve.
The results are then shown in a plot of the peak position (i.e. a slope
value) versus the length of the portion, with the peak widths as error
bars (see Fig. \ref{fig:2}b).
If the analyzed curve presents a linear region, this plot shows a plateau
for $\ell$ ranging from $\ell_{min}$ to the length of the whole linear
region. This plateau is usually very easy to be identified because of the
distinction between linear and curved portions. In fact, portions of
length larger than the length of the whole linear region are considered
curved portions and discarded. Thus, the plot usually drops to zero at the
end of the plateau. Eventually, through an average and a standard
deviation, we obtain the final slope value and its fitting error, while
the length of the plateau gives the length of the linear region. In
conclusion, our AFP is able to identify not only the slope of the linear
region but also its length. We have tested our AFP before its application
to the systematic analysis and we have found that the measured Hurst
exponent is widely independent of the fitting parameter \footnote{As
Hamburger \textit{et al.} did in Ref. \cite{ham96}, we have not chosen a
value of the parameter, being this choice not relevant to the purposes of
the article. Nevertheless, we have tested a wide range of reasonable
values of the parameter in order to be sure of the repeatability of the
results obtained.}. Conversely, the length of the linear region strongly
depends upon the value of the parameter and must be considered only an
internal parameter of the analysis and not a direct measurement of the
scale invariance range.

\section{Numerical Analysis}

With all the generation algorithms published in literature we have created
sampled self-affine profiles with known fractal dimension $D=2-H$. We have
varied the exponent $H$ between 0.1 and 1 and we have focused on the value
$L=512$ sampling points (the best sampling obtainable with most of the
SPMs). We discuss also different values of $L$ up to 16384. Because there
exists only a few algorithms that generate exactly self-affine profiles,
we have used algorithms that generate statistically self-affine profiles,
which are more difficult to handle but closer to reproduce natural
physical systems. The algorithms we have used are known in literature as:
the random midpoint displacement \cite{sch95,vos86a}, the random addition
algorithm \cite{fed88,pei89}, the fractional Brownian motion \cite{fed88},
the Weierstrass-Mandelbrot function \cite{lop94,ber80}, the inverse
Fourier transform method \cite{vos86a} and a variation of the independent
cut method \cite{fal90}. For the measurement of the Hurst exponent of
self-affine profiles we have used the height-height correlation function
$C_2$ \cite{yan97} and the root mean square variable bandwidth with fit
subtraction method \cite{mor94,pen94}. The value of $H_{out}$ has been
calculated from the slope in the log-log plot of the average over $N$
statistically independent AFs, measured with our AFP.

The results are expressed in terms of $H_{out}$ vs. $H_{in}$ plots. Each
plot is characteristic of a single AF and generation algorithm and it
represents the relationship between the measured Hurst exponent $H_{out}$,
calculated from the average of $N$ AFs, and the nominal exponent $H_{in}$
of the profile. Grouping the $H_{out}$ vs. $H_{in}$ plots obtained using
the same AF for all the generation algorithms, the dispersion of the
$H_{out}$ values comes to evidence.
\begin{figure}
\begin{center}
\includegraphics{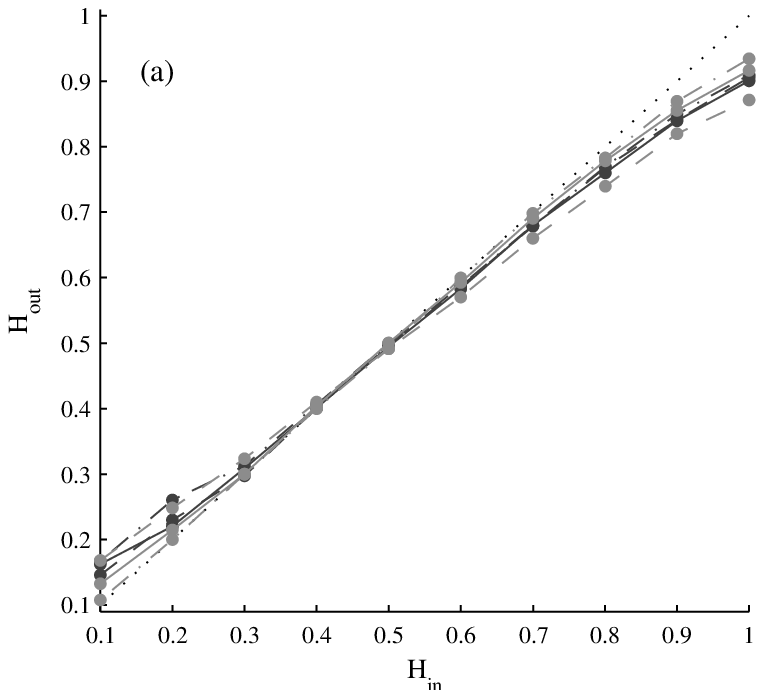}
\includegraphics{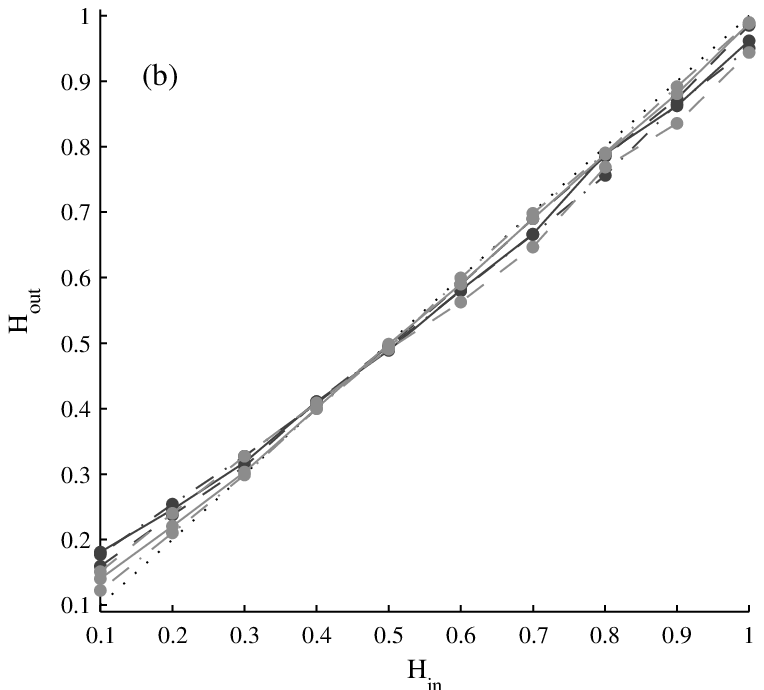}
\caption{\label{fig:3}$H_{out}$ vs. $H_{in}$ graphs calculated from N =
500 profiles of L = 512 points each: (a) Height-height correlation
function and (b) Root mean square variable bandwidth (with fit
subtraction). The black dotted line represents the ideal $H_{out} =
H_{in}$ behavior. The other line styles are related to different
generation algorithms: random midpoint displacement (black continuous
line), inverse Fourier transform (black dashed line), random addition
(black dash-dotted line), Weierstrass-Mandelbrot (grey continuous line),
fractional Brownian motion (grey dashed line) and independent cut (grey
dash-dotted line).}

\end{center}
\end{figure}
In Fig. \ref{fig:3} we show the $H_{out}$ vs. $H_{in}$ graphs obtained
from $N=500$, $L=512$ profiles, as explained in the previous section. We
show separately in Figs. \ref{fig:3}(a) and \ref{fig:3}(b) the different
AFs used. Since the profiles are \textit{statistically} self-affine, the
measured $H_{out}$ are subject to a statistical error that is inversely
related to $N$ \cite{den99b}. In order to characterize the dependence of
this statistical error on the number $N$ of averaged AFs, we let $N$ vary
from 1 to 50 using the same profiles considered in Fig. \ref{fig:3}. With
these values of $N$ we have repeated the numerical analysis (i.e.
calculation of the AFs, averaging and application of the AFP) and we have
extracted a standard deviation $\sigma_{N}$ of the measured exponents.
\begin{figure}
\begin{center}
\includegraphics{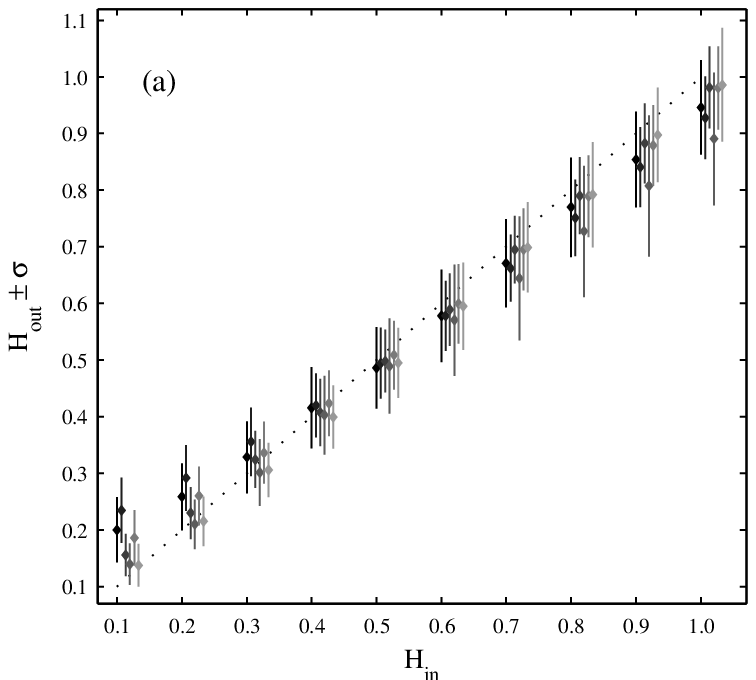}
\includegraphics{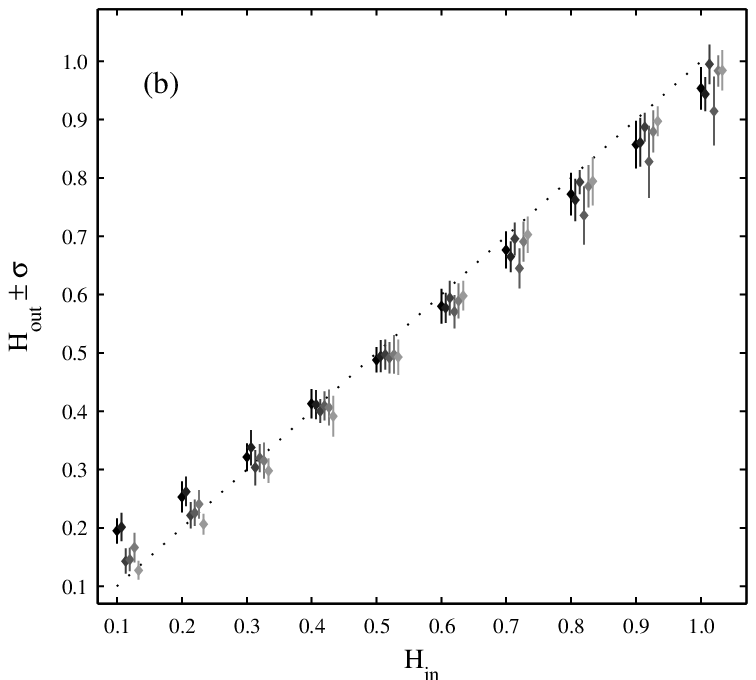}
\includegraphics{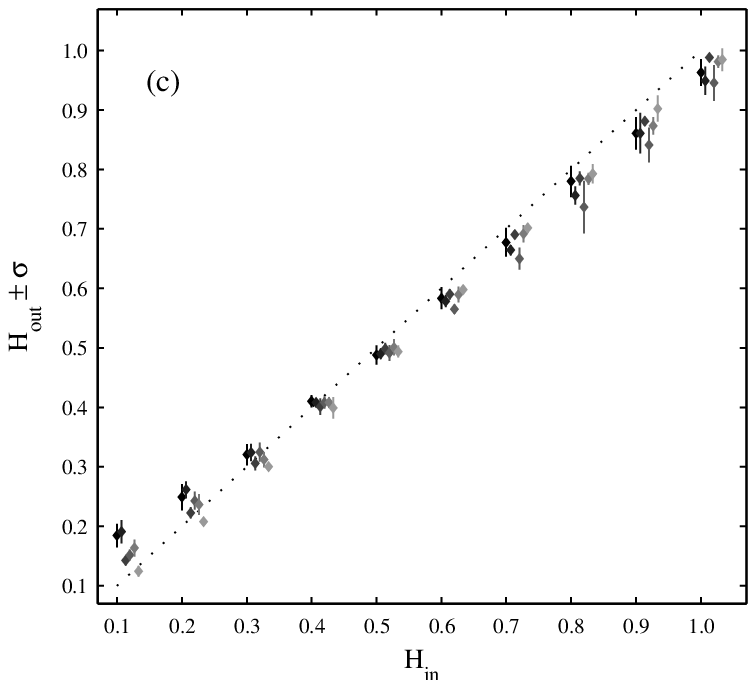}
\caption{\label{fig:4}$H_{out}$ vs. $H_{in}$ graphs with error bars equal
to twice the standard deviation $\sigma_N$ of the measured exponents.
These graphs correspond to different values of the number $N$ of
statistically independent profiles from which an average Hurst exponent is
measured: (a) $N=1$, (b) $N=10$ and (c) $N=50$. It can be seen that for $N
> 10$ and for $H_{in} < 0.3$ the overlap between the error bars
corresponding to different generation algorithms is small or completely
absent. For the sake of clarity we do not distinguish between different
generation algorithms}.
\end{center}
\end{figure}
In Fig. \ref{fig:4} we show the $H_{out}$ vs. $H_{in}$ graphs, analogous
to those in Fig. \ref{fig:3}, with the calculated error bars (twice the
standard deviation $\sigma_{N}$), for a few values of $N$. We present the
results for a single AF (the root mean square variable bandwidth with fit
subtraction), the results for the other AFs being similar.
\begin{figure}
\begin{center}
\includegraphics{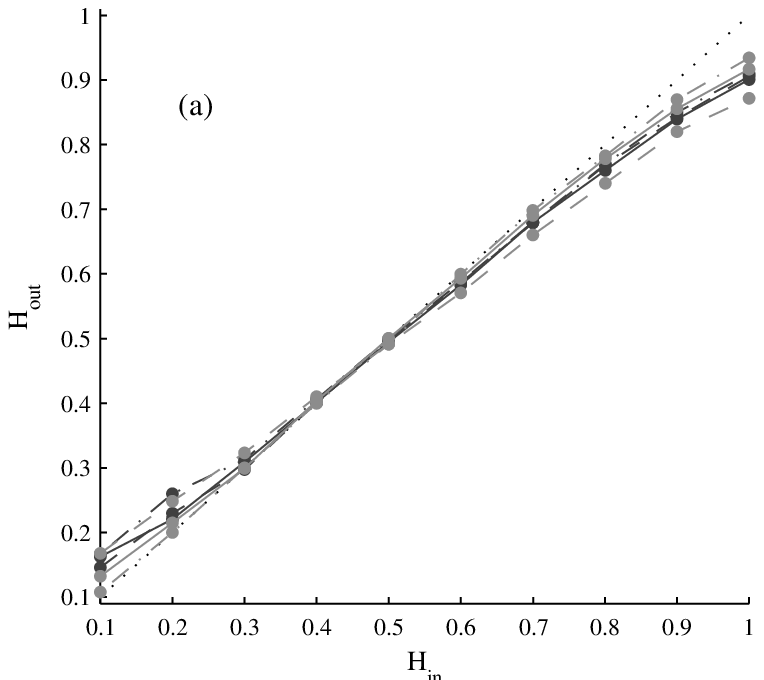}
\includegraphics{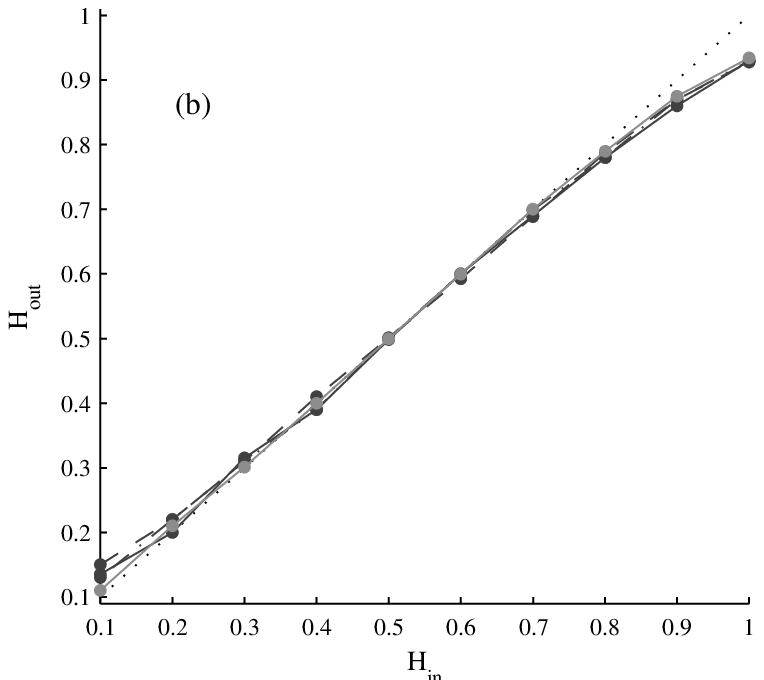}
\includegraphics{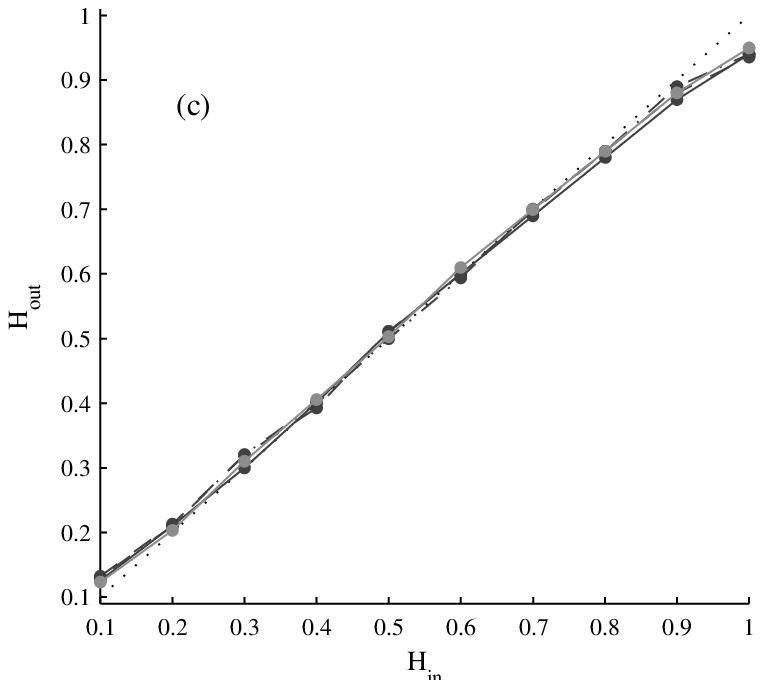}
\caption{\label{fig:5}$H_{out}$ vs. $H_{in}$ graphs calculated with the
height-height correlation function from: (a) $N = 500$, $L = 512$
profiles. (b) $N = 50$, $L = 4096$ profiles. (c) $N = 15$, $L = 16384$
profiles. Line styles are the same as in Fig. \ref{fig:3}.}
\end{center}
\end{figure}
In Fig. \ref{fig:5} we show three $H_{out}$ vs. $H_{in}$ graphs obtained
respectively with $N=500$, $L=512$ profiles, $N=50$, $L=4096$ profiles and
$N=15$, $L=16384$ profiles. Again, we present only one AF (the
height-height correlation function $C_2$).

\section{\label{sec:results&discussion}Results and Discussion: Deviation and Dispersion from the Ideal
Behavior}

Ideal continuous fractal profiles are statistically characterized by their
fractal dimension (universality) and their $H_{out}$ vs. $H_{in}$ graphs
are straight lines \cite{bar95,fed88,fal90}.

In Fig. \ref{fig:3} a deviation from the ideal behavior is observed for
both the AFs. It turns out that the sampling of a profile affects in a
different way different methods of analysis. The deviation from the ideal
behavior has been already observed in literature (for example, see Ref.
\cite{sch95}) and our results are in good agreement with the previous
ones.

Moreover, within the same method of analysis we observe that the different
generation algorithms give significantly different $H_{out}$ vs. $H_{in}$
plots. This dispersion is pointed out here for the first time because
different generation algorithms are considered together. The significance
of the dispersion can be inferred from the characterization of the
statistical error of the measured exponent discussed hereafter.

In Fig. \ref{fig:4} we show that for $N>25$ and $H_{out}<0.3$ the error
bars of $H_{out}$ for different generation algorithms hardly overlap. This
fact suggests that the statistical error is not the only reason of the
differences between the $H_{out}$ vs. $H_{in}$ plots shown in Fig. 3.
\begin{figure}
\begin{center}
\includegraphics{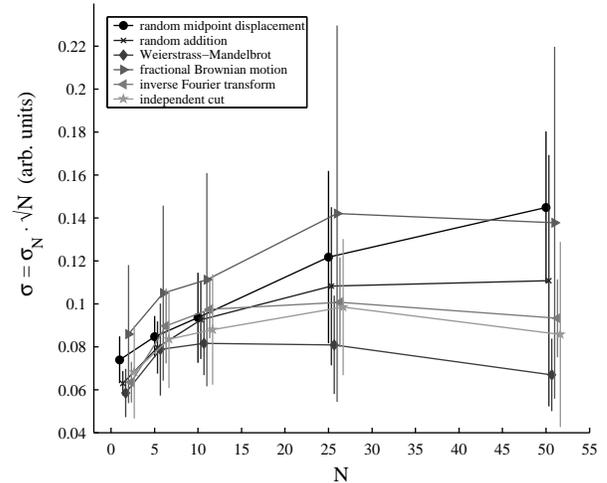}
\caption{\label{fig:6}Graph of the statistical standard deviation $\sigma$
of the Hurst exponent, obtained from the definition of the standard
deviation of the mean (Eq. (\ref{eq:sigma})), vs. the number $N$ of
averaged AFs. It can be clearly seen the saturation for values of $N$
bigger than 25 for almost all the generation algorithms.}
\end{center}
\end{figure}
In Fig. \ref{fig:6} we plot the statistical error $\sigma_{N}$ times the
square root of $N$ vs. $N$. For $N\geq10$ the curves approach a constant
value according to the relationship between the standard deviation of
independent, normally distributed measurements and the standard deviation
of the mean upon $N$ measurements:
\begin{equation}
\sigma_{N}=\frac{\sigma}{\sqrt{N}}  \label{eq:sigma}
\end{equation}
This result shows that the AFP and the averaging of the AFs do commute.
The assessment of this property is non-trivial due to the complexity of
the AFP. Thus, we extrapolate the statistical error of the measured
exponents in Fig. \ref{fig:3} ($N=500$) using Eq. (\ref{eq:sigma}) where
$\sigma$ is extracted from the plateau in Fig. \ref{fig:6}. Overestimating
$\sigma$ with the value 0.16 we obtain $\sigma_{500}=0.007$. This value
produces an error bar in Fig. \ref{fig:3} as small as the symbol used to
mark the data. A direct calculation of $\sigma_{500}$, obtained averaging
AFs calculated on groups of $N=500$ profiles for every $H\in [0.1,1]$ and
for every generation algorithm, fitting and extracting a mean value and a
standard deviation of $H$, would have required a huge and time consuming
calculation.

These results suggest that the observed dispersion between the $H_{out}$
vs. $H_{in}$ curves for different generation algorithms is an intrinsic
effect of the sampling, depending only on the number of sampling points
$L$. This fact has an important consequence on a fractal analysis of
experimental surfaces. While looking at a real sample, we do not know what
kind of ``algorithm'' has generated the surface. This introduces an
uncertainty on its real fractal dimension independent of the statistical
error. Thus, there is an intrinsic upper limit to the precision of the
measurement of the exponent. It is useless to strengthen the statistics
once the number of acquired profiles makes the statistical error smaller
than the intrinsic dispersion.

In Fig. \ref{fig:5} we see that as $L$ increases both the deviation and
the dispersion decrease in agreement with their expected vanishing in the
limit of $L$ going to infinity \cite{sch95}. This is also an \textit{a
posteriori} proof of the correctness of both the generation algorithms and
the methods of analysis.

Our interpretation of these effects is that the sampling of a self-affine
profile lessens its fractality in such a way that it is no longer
characterized universally by its fractal dimension (or Hurst exponent).
While for a continuous self-affine profile the relationship
$H_{out}=H_{in}$ holds, for sampled profiles we can see that different AFs
produce different $H_{out}$ vs. $H_{in}$ plots from the same sampled
fractal profile. Considering instead a single AF, our results show that
sampled fractal profiles generated with different generation algorithms
but with the same ideal dimension give different measured Hurst exponents.

However, Fig. \ref{fig:5} clearly shows that the lessening of fractality
of a profile is rather a continuous process than a sharp transition: the
poorer is the sampling, the worse are the deviation and the dispersion. In
Figs. 5 and 3 we observe that the lessening of fractality acts in a
similar way on profiles generated with different algorithms. The common
trend of the $H_{out}$ vs. $H_{in}$ curves obtained from different
generation algorithms is interpreted as a consequence of the universality
of fractal objects.

It is then reasonable to assume the existence for every AF of a universal
region in the $H_{out}$-$H_{in}$ plane containing all the $H_{out}$ vs.
$H_{in}$ plots obtained with every possible generation algorithm. This
region, approximately identifiable with the envelope of the $H_{out}$ vs.
$H_{in}$ plots, has a width that depends on the number of sampling points
and approaches the 1-dimensional $H_{out}=H_{in}$ ideal curve for very
large values of $L$. We expect that, given any continuous self-affine
profile with a Hurst exponent $H_{in}$ and given the exponent $H_{out}$
measured from an $L$-point sampling of the continuous profile, the pair
($H_{in}$,$H_{out}$) belongs to the universal region of the corresponding
graph (specific for every AF and number of sampling points $L$). Provided
a good characterization of the aforementioned regions (i.e. using as many
generation algorithms as possible), we can use them to generate
calibration graphs for every $L$ and AF describing the relationship
between the measured $H_{out}$ and the true value $H_{in}$.

\begin{figure}
\begin{center}
\includegraphics{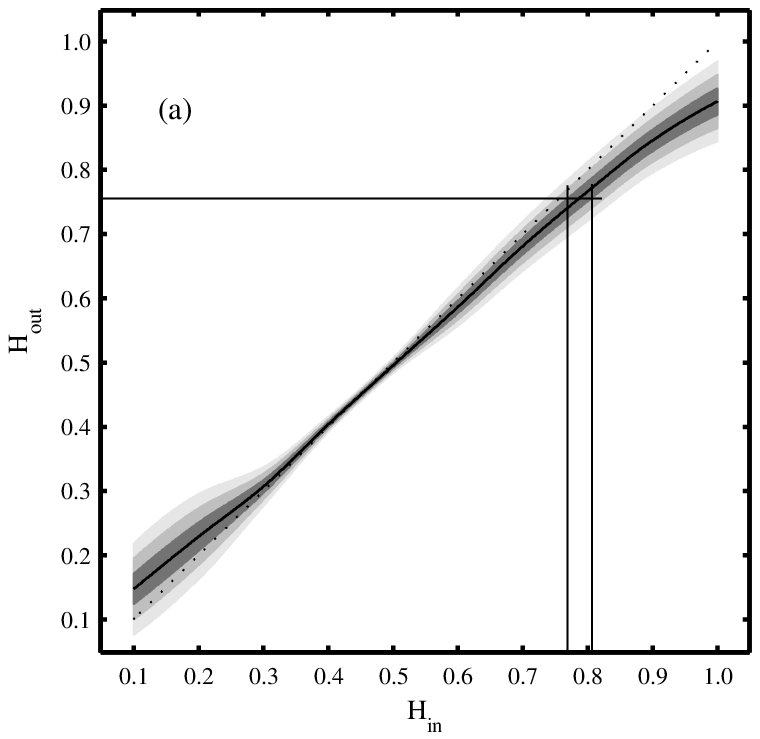}
\includegraphics{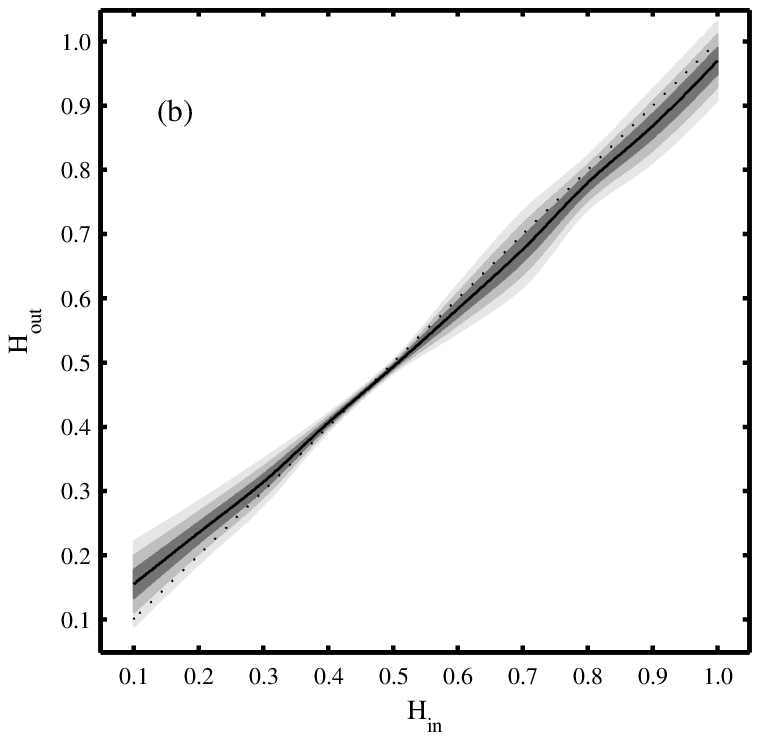}
\caption{\label{fig:7}Calibration graphs $H_{out}$ vs. $H_{in}$ for the
methods of analysis used in this article: (a) height-height correlation
function and (b) variable bandwidth with fit subtraction. From the value
of the measured exponent, one can easily extract the corresponding
confidence interval of the corrected exponent, as represented graphically
in (a).}
\end{center}
\end{figure}

To produce the calibration graphs we proceed as follows. First of all, we
make two general assumptions in order to take quantitatively into account
the problem of measuring the Hurst exponent of a sampled profile. We
assume that the $H_{out}$ values corresponding to the same $H_{in}$ are
normally distributed around a mean $\langle H_{out}\rangle$, and we assume
also that the values obtained with the available generation algorithms are
a random sampling of the gaussian distribution. We then measure the
average and the standard deviation of the dispersed $H_{out}$ values
corresponding to each $H_{in}$ separately. Thus we obtain a sampling of
the functions describing the dependence of $\langle H_{out}\rangle$ and
$\sigma_{H_{out}}$ from $H_{in}$. With an interpolation algorithm using
smooth functions, we derive the curve representing the relationship
between $\langle H_{out}\rangle$ and $H_{in}$. We also derive the pair of
curves corresponding to $\langle H_{out}\rangle+n\sigma_{H_{out}}$ and
$\langle H_{out}\rangle-n\sigma_{H_{out}}$ vs. $H_{out}$ which define the
$n$-th confidence level. For every value of $H_{out}$ it is possible to
find the confidence interval of $H_{in}$ for any given confidence level.
The resulting graphs for $L=512$ are shown in Fig. \ref{fig:7}.
These calibration graphs allow to take into account the deviation and the
dispersion due to the sampling. A similar method has been independently
proposed in Ref. \cite{den99a} even though the analysis was limited to a
single generation algorithm and the discussion on the reliability of the
calibration regions together with the intrinsic dispersion were completely
neglected.

Using the calibration graphs it is possible to measure the Hurst exponent
of poorly sampled profiles correcting for the first time the deviation due
to the sampling and providing a reasonable estimate of the error on a
confidence level basis. The quantification of the error is of paramount
importance, as pointed out in the introduction, since many authors
estimated the error from the precision of the linear fit
\cite{kri93,iwa93} or from the standard deviation of the measured
exponents \cite{den99b}. Our results show that they usually underestimated
the true error.

\section{\label{sec:DP}Application of the calibration graphs to the study of directed
percolation numerical profiles}

We have applied our procedure to the 1+1 dimensional directed percolation
(DP) model, described by S.V. Buldyrev \textit{et al.} \cite{bul92}. This
model mimics the paper wetting process by a fluid. The resulting pinned
interface is self-affine with exponent $H\simeq 0.63$.

We have analyzed $N=30$, $L=16384$ DP profiles with the height-height
correlation function (\textit{h.-h. corr}) and the variable bandwidth with
fit subtraction (\textit{vbw}), using the automated fitting protocol to
measure the Hurst exponents. The results are shown in the second column of
Tab. \ref{tab:DP}. We have not calculated the statistical error (see
Section \ref{sec:results&discussion}) because it would have been
excessively time consuming. Thus, the error shown is simply the error of
the fit calculated with the AFP.
\begin{table*}
\caption{\label{tab:DP}Measured Hurst exponents of sampled DP profiles
(theoretical value: $H\simeq 0.63$ \cite{bul92})}  
\begin{ruledtabular}
\begin{tabular}{|c|c|c|c|}
   & \hspace{1cm}$H_{out}^{16384}$\,
   \footnote{The error for $L=16384$ is the error of the fit.}
   \hspace{1cm} & \hspace{1cm}$H_{out}^{512}$\,
   \footnote{The error for $L=512$ is the \textit{rms} value of the statistical error and the error of the fit}
   \hspace{1cm} & \hspace{1cm}$H_{in}^{512}\;(68\%)$\hspace{1cm} \\
\hline
\hline
\hspace{1cm}\textit{h.-h. corr.}\hspace{1cm}          & $0.615\pm0.004$     & $0.609\pm0.002$    & $[0.613-0.635]$\\
\hline
\textit{vbw}       & $0.620\pm0.003$     & $0.608\pm0.012$    & $[0.611-0.644]$\\
\end{tabular}
\end{ruledtabular}
\end{table*}
The values of the measured exponents $H_{out}^{16384}$ are significantly
lower than the ones predicted by the DP model, suggesting that a
correction is needed even in the case of profiles of $L=16384$ points,
which are widely considered as continuous.

We have then analyzed $N=1000$, $L=512$ profiles extracted from the
$L=16384$ profiles. We have applied the correction procedure based on the
calibration graphs shown in Fig. \ref{fig:7} to the exponents measured
with the AFP. In the third column of Tab. \ref{tab:DP}, the uncorrected
measured exponents ($H_{out}^{512}$) are shown. The error is calculated as
the root mean square (\textit{rms}) value of the statistical error
$\sigma_{1000}$ (evaluated as explained in Section
\ref{sec:results&discussion}) and the error of the fit calculated with the
AFP. In the fourth column, the confidence intervals corresponding to the
$68\%$ probability for the ``true'' exponents are shown
($H_{in}^{512}\;(68\%)$).

The results summarized in Tab. \ref{tab:DP} allow to notice the
effectiveness of the calibration graphs in the analysis of self-affine
profiles when the effects of sampling are non negligible. In the example
reported here, the poor sampling causes a discrepancy of about $4\%$
between the measured exponents and the theoretical one for DP profiles.
After the correction with the calibration graphs, the expected value
$H_{in}\simeq 0.63$ is consistent with the confidence intervals of the
three AFs. Moreover, the intrinsic error due to the dispersion (about half
the width of the confidence interval) turns out to be usually one order of
magnitude larger than the aforementioned \textit{rms} error.

In conclusion, our calibration graphs have allowed to correct the
deviation and to quantify the intrinsic error of the Hurst exponent of
poorly sampled ($L=512$) DP profiles.

\section{Conclusions}

We have carried out a systematic analysis in order to achieve a deeper
understanding of the effects of sampling on the measurement of the Hurst
exponent of self-affine profiles. This is a crucial point for the
assessment of the reliability of fractal analysis of experimental
profiles, such as topographic profiles of growing thin films and
interfaces acquired with a Scanning Probe Microscope. We have pointed out
that some of the steps leading to the measurement of the Hurst exponent
have been only superficially discussed, although worth of deeper
attention. We have focused on the quantification of the effects of
sampling and possibly on their correction, allowing a more reliable
identification of the universality class of growth.

In order to perform such a quantitative analysis we have developed a new
automated fitting protocol that allows to remove the ambiguity in the
choice of the region for the linear fit of the analysis functions. This
point is usually underestimated in the published experimental literature,
and appears to be a significant source of error in the whole analysis.
Moreover, an automated protocol sensibly reduces the time required for the
fitting of a large number of noisy curves, allowing a higher statistics.
With our automated fitting protocol we have systematically investigated
synthetic self-affine profiles generated with all the generation
algorithms found in literature using different method of analysis.

The systematic analysis presented in this paper has been carried out on
1+1 dimensional profiles and we have not considered 2-dimensional methods
of analysis (e.g. see \cite{den99a,kri93}). However, it is reasonable to
suppose that even in this case the effects of sampling cannot be
neglected, and the conclusions drawn in Ref. \cite{den99a} are probably
incorrect. The similarity between Fig. 1 in Ref. \cite{den99a} and the
analogous results presented in this paper (see the variable bandwidth
analysis of profiles generated with the random midpoint displacement shown
in Fig. \ref{fig:3}c) suggests that conclusions very close to those
presented here can be drawn also in the 2-dimensional case.

Studying the discrepancy between the measured Hurst exponent $H_{out}$ and
the ``true'' one ($H_{in}$) for synthetic self-affine profiles with
$L=512$ points, we have shown that the main effects of sampling are a
deviation of the $H_{out}$ vs. $H_{in}$ plots from the ideal behavior and
a dispersion of the exponents calculated from different generation
algorithms. Both these effects smoothly reduce with increasing values of
$L$. The deviation turns out to be universal in the sense that the trend
of the $H_{out}$ vs. $H_{in}$ curves is common to all of the generation
algorithms, depending only on the number of sampling points and on the
function used in the analysis. We propose that this behavior is
reminiscent of the fact that a fractal object is completely characterized
by its dimension and therefore the deviation can be at least empirically
corrected. The dispersion instead has to be considered as an intrinsic
error due to the sampling, but for the very special case of profiles whose
generation algorithm allows to build their specific $H_{out}$ vs. $H_{in}$
plot. This dispersion error must be quantitatively taken into account
since it cannot be reduced with an increase in the statistics but only
with an increase in the number of sampling points.

The existence of an intrinsic dispersion error in the measurement of the
Hurst exponent that depends only on the number of sampling points is very
important. In fact, this intrinsic error easily overwhelms the statistical
error for poorly sampled profiles. It is definitely clear that a reliable
result cannot be based on the consideration of the statistical error only.
Moreover, the dispersion poses an upper limit to the precision in the
measurement of the Hurst exponent of sampled profiles. It becomes useless
to increase the statistics once the statistical error has been made
reasonably smaller than the intrinsic one. This is particularly important
in an experimental analysis because it usually reduces significantly the
number of profiles that have to be acquired, making the analysis much less
time consuming.

Thanks to our systematic analysis, we have built, for each method of
analysis, a calibration graph representing the region of the
$H_{out}$-$H_{in}$ plane where the true exponents fall within a given
confidence level. We have originally proposed to use these graphs as a
reliable empirical method to correct the measured value of the Hurst
exponent of a poorly sampled profile and to estimate its intrinsic
sampling error. The reliability of the calibration graphs is based on two
assumptions:
\begin{enumerate}
\item[i)] The measured exponents for all the possible self-affine profiles, with
the same ``true'' exponent $H_{in}$ and with the same number of sampling
points, are normally distributed;
\item[ii)] The numerical generation algorithms known in literature provide a
statistically reliable sample of all the possible self-affine profiles.
\end{enumerate}
Even though we have found just six generation algorithms in literature, we
believe that they still allow to obtain reasonable results or at least the
only ones obtainable to date. These results represent a step forward to a
reliable fractal analysis of both numerical and experimental profiles and
to the individuation of the universality classes in the study of the
evolution of many different systems.

In conclusion, we have demonstrated that a reliable measurement of the
Hurst exponent of poorly sampled self-affine profiles is possible,
provided that the measured $H_{out}$ is corrected of its deviation and
that the sampling error is quantitatively taken into account. We have thus
given strength to experimental analyses, since the numerical results
reported in literature to date led to the conclusion that the analysis of
self-affine profiles sampled with less than 1000 points is not reliable
\cite{sch95}. Even with the great improvement introduced by the use of the
calibration graphs in the analysis of self-affine profiles, we definitely
agree with Schmittbuhl \textit{et al.} in pointing out that the comparison
of the results obtained with different method of analysis is of
fundamental importance \cite{sch95}. Furthermore, we shortly comment on
the common experimental procedure of connecting AFs calculated from
profiles acquired with different scan sizes \cite{kri93,iwa93,fan97}. This
connection allows investigating a wider range of length scales with a
limited number of sampling points and makes the measurement more reliable.
However, the deviation and dispersion are not influenced by this
procedure, since they depend only on the number of sampling points of the
profiles on which the AFs are calculated.

The AFP and the calibration graphs have been tested on numerically
generated 1+1 dimensional directed percolation (DP) profiles, which have
provided a benchmark to check our protocol. We have shown that for $L=512$
profiles a correction is needed and the calibration graphs allow to
recover the theoretical value of $H$ predicted by the DP model. We have
also shown that a correction is needed even for the $L=16384$ profiles,
which are widely considered as continuous.

Our results provide a powerful tool for the accurate extraction of the
Hurst exponent from poorly sampled profiles, and for the quantification of
the error in the measurement. This is of paramount importance for
experimentalists who study the scale invariance of surfaces and interfaces
by Scanning Probe Microscopy or other techniques, with the aim of
identifying the underlying universality classes. The huge amount of
experimental results published in the past two decades about the
fractality of many interfaces can be now analyzed under a new light.

\begin{acknowledgments}
We thank E.H. Roman and G. Benedek for discussions. Financial support from
MURST under the project COFIN99 is acknowledged.
\end{acknowledgments}

\end{document}